\documentclass[reqno]{amsart}
\usepackage[english]{babel}
\usepackage[latin1]{inputenc}
\usepackage{latexsym,amsfonts}
\usepackage{graphics}
\usepackage{ulem} \usepackage{hhline}
\usepackage{dsfont} \usepackage{mathrsfs}
\usepackage{mathpazo} \usepackage{euscript} \usepackage[pdftex]{hyperref}
\makeatletter\@addtoreset{equation}{section}

\newtheorem {theorem}{Theorem}[section]
\newtheorem {definition}[theorem]{Definition}
\newtheorem {remark}[theorem]{Remark}
\newtheorem {proposition}[theorem]{Proposition}

\newcommand{\C}{\mathbb C} \newcommand{\R}{\mathbb R}  \newcommand{\D}{\mathbb D}
\newcommand{\B}{\mathscr B} \newcommand{\T}{\mathscr T} \newcommand{\W}{\mathscr W} 
\newcommand{\Lap}{\mathscr L} \newcommand{\sH}{H} \newcommand{\bet}{2\nu}
\newcommand{\mes}{d\lambda}

\begin{document}
\title[Generalized second Bargmann transforms]{Generalized second Bargmann transforms associated
with the hyperbolic Landau levels on the Poincaré disk }
\author[ElWassouli, Ghanmi, Intissar, Mouayn]{ Fouzia ElWassouli, Allal Ghanmi, Ahmed Intissar and Zouha\"{i}r Mouayn}
\address{{\bf Fouzia ElWassouli,} \quad Department of Mathematics, Faculty of  Sciences,
   Ibn Tofa\"{i}l University, Kénitra - Morocco}
\address{{\bf Allal Ghanmi $\&$  Ahmed Intissar} \quad  Department of Mathematics,  Faculty of  Sciences, P.O. Box 1014,
   Mohammed V University,  Agdal,  10 000 Rabat - Morocco  }
\address{{\bf Zouha\"{i}r Mouayn,} \quad Department of Mathematics, Faculty of Sciences and Technics (M'Ghila), P.O.
Box 523, Sultan Moulay Slimane University, 23 000 B\'{e}ni Mellal - Morocco}

\begin{abstract}
We deal with a family of generalized coherent states associated to the
hyperbolic Landau levels of the Schrödinger operator with uniform magnetic
field on the Poincaré disk. Their associated coherent state transforms
 constitute a class of generalized second Bargmann transforms.
\end{abstract}
\maketitle
\section{Introduction}

The classical Bargmann transform made the space of square integrable functions $f$ on the real
line isometric to the space of entire functions that are
$e^{-|z|^{2}}d\mu$-square integrable, $d\mu $ being the Lebesgue measure
on the complex plane. It is defined in \cite{Bargmann61} as
\begin{eqnarray}
 \B[f](z):=\pi ^{-\frac{1}{4}}\int_{\R}\exp \left( -\frac{\xi^{2}}{2}+\sqrt{2}\xi z-\frac{z^{2}}{2}\right) f(\xi )d\xi ;
           \qquad z\in \C,  \label{1.1}
\end{eqnarray}
where the involved kernel function is related to the generating function of the Hermite polynomials.

In the same paper \cite[p.203]{Bargmann61}, V. Bargmann has also introduced a second transform $\T_{\nu}$; $\nu>1/2$, as a unitary
integral operator whose kernel function corresponds to the generating function of the Laguerre polynomials $L_m^{(\alpha)}(\cdot)$ \cite{Rainville,Szego}. It maps isometrically the space of $\xi^{\bet}d\xi/\Gamma(\bet)\xi$-square integrable functions on the positive real half-line onto the
weighted Bergman space
\begin{eqnarray}   \label{WBS}
 \mathcal{A}^{2,\nu }(\D):=\left\{ f\mbox{ \, holomorphic on }\,%
\D;\int_{\D}|f(z)|^{2}\left( 1-|z|^{2}\right)^{\bet-2}d\mu (z)<\infty \right\}
\end{eqnarray}
on the unit disk $\D=\left\{ z\in \Bbb{C};\,|z|<1\right\} $. Explicitly, we have
\begin{eqnarray}
 \T_{\nu}[\psi](z)=\left(\frac{{\bet-1}}{\pi}\right)^{1/2}
(1-z)^{-\bet}\int_{0}^{\infty}\exp\left( {-\frac{\xi }{2}\left( \frac{1+z}{1-z}\right) }\right){\psi (\xi )} \xi^{\bet}(\Gamma(\bet))^{-1}\frac{d\xi}{\xi},
\end{eqnarray}
where the positive real number $\bet-1$ represents the parameter $\gamma$ in \cite[p.203]{Bargmann61}.
Thus, using the canonical isometry from $L^{2}(\R_{+}^{*},d\xi/\xi)$ onto $L^{2}\left(\R_{+}^{*},\xi^{\bet}d\xi/\Gamma(\bet)\xi \right)$,
one extends $\T_{\nu}$ to the transform
\begin{eqnarray}
\W_{\nu}[\varphi ](z) 
:=\left(\frac{\bet-1}{\pi\Gamma(\bet)}\right)^{1/2} (1-z)^{-\bet}
\int_{0}^{+\infty} \xi^{\nu} \exp\left(-\frac{\xi }{2}\left( \frac{1+z}{1-z}\right) \right) {\varphi (\xi)}\frac{d\xi}{\xi}   \label{SBTM}
\end{eqnarray}
 mapping isometrically the Hilbert space $L^{2}(\R_{+}^{*},d\xi/\xi )$ onto $\mathcal{A}^{2,\nu}(\D)$.
Note that the space $\mathcal{A}^{2,\nu}(\D)$ in \eqref{WBS} can also be realized as the null space of the second order differential operator
\begin{eqnarray}
{\sH}_{\nu} = -4(1-|z|^{2})\left( (1-|z|^{2})\frac{\partial ^{2}}{\partial z\partial \overline{z}}
- \bet  \bar z\frac{\partial}{\partial \bar z} \right). \label{zhangOP}
\end{eqnarray}
The latter one is acting on the Hilbert space $L^{2,\nu}(\D):=L^{2 }(\D,(1-|z|^{2})^{\bet-2}d\mu)$ and can be unitarily intertwined to represent an Hamiltonian of uniform magnetic field on the unit disk.

In this paper, we will be concerned with the $L^2$-eigenspaces
 \begin{eqnarray}
\mathcal{A}_{m}^{2,\nu}(\D):=\left\{ F\in L^{2,\nu}(\D);\qquad {\sH}_{\nu}F=\epsilon _{m}^{\nu}F \right\}
\label{GBS}
\end{eqnarray}
associated to the discrete spectrum of ${\sH}_{\nu}$ consisting of the eigenvalues ({\it hyperbolic Landau levels}):
\begin{eqnarray}
\epsilon _{m}^{\nu}= 4m(\bet-m-1)  ; \qquad m=0,1,2, \cdots , [\nu-(1/2)], \label{DiscSpec}
\end{eqnarray}
where $[x]$ denotes the greatest integer less than $x$.
 Our aim is to construct a family of integral transforms generalizing \eqref{SBTM}. The method used is
  on the coherent states analysis  ``\textit{à la Iwata}'' \cite{Iwata} together with the concrete description of the $L^{2}$-spectral theory of
 the operator $\sH_{\nu}$ \cite{elstrodt,fay,zhang}. Precisely, we establish the following

\begin{theorem} \label{Thm}
Let $\nu$ be a real number such that $\nu>1/2$ and  $m=0,1,2,\cdots, [\nu-(1/2)]$. Then, the mapping
\begin{align}
\W_{\nu ,m}[\varphi ](z)&= \left( \frac{(2(\nu-m)-1)m!}{\pi\Gamma(\bet-m)}\right)^{\frac{1}{2}}
\left( \frac{|1-z|}{1-|z|^{2}}\right) ^{2m}(1-z)^{-\bet} \label{GBerInt}\\
& \times \int_{0}^{+\infty}\xi^{\nu-m}
\exp \left( -\frac{\xi }{2}\left( \frac{1+z}{1-z}\right) \right)
 L_{m}^{2(\nu-m)-1}\left( \xi \frac{(1-|z|^{2})}{|1-z|^{2}}\right) {\varphi (\xi)}\frac{d\xi}{\xi}\nonumber
\end{align}
 defines a unitary isomorphism from $L^{2}(\R_{+}^{*},d\xi/\xi )$ onto the Hilbert space
$\mathcal{A}_{m}^{2,\nu}(\D)$.
The particular case of $m=0$ reproduces to the second Bargmann transform $\W_{\nu}$ in \eqref{SBTM}.
\end{theorem}

The paper is organized as follows. In Section 2, we review some needed facts
on the $L^{2}$-spectral theory of the operator ${\sH}_{\nu}$
on the unit disk. In Section 3, we recall the coherent states
formalism we will be using. In Section 4, we define a family of generalized
coherent states attached to hyperbolic Landau levels. The
associated coherent state transforms constitute to a family of generalized second
Bargmann transforms.

\section{Spectral analysis of ${\sH}_{\nu}$; $\nu>0$}

The second order differential operator $\sH_{\nu}$ in \eqref{zhangOP}
appears as the Laplace-Beltrami operator on the unit disk perturbed by a first order differential operator.
 It can be interpreted as the Hamiltonian of a charged particle moving in an external uniform magnetic
field. In fact, $\sH_{\nu}$ is unitary equivalent to the magnetic Schrödinger operator \cite{GhInJMP}:
\begin{eqnarray}
\Lap_{\nu}:=(d+\sqrt{-1}\nu \theta )^{*}(d+\sqrt{-1}\nu \theta ),
\label{2.4}
\end{eqnarray}
 associated to the gauge potential vector $\theta (z)=-\sqrt{-1}(\partial -\bar{\partial}){\log }(1-|z|^{2})$,
 and acting on $L^{2}(\D)=L^{2}(\D,(1-|z|^2)^{-2}d\mu)$. Indeed, we have
$(1-|z|^{2})^{\nu }{\sH}_{\nu}(1-|z|^{2})^{-\nu }=\Lap_{\nu}.$
  Different aspects of its spectral analysis have been studied by many authors,
   e.g. \cite{elstrodt,fay,zhang}. For instance, note that ${\sH}_{\nu}$ is
  an elliptic densely defined operator on the Hilbert space $L^{2,\nu}(\D)$ and
  admits a unique self-adjoint realization that we denote also by ${\sH}_{\nu}$.
  Note also that such operator commutes with the action of the group $SU(1,1)$ 
  defined on $L^{2,\nu}(\D)$ by
$$(T^{\nu}_gf)(z) :=  (\det(g')(z))^\nu  f(g.z); 
\quad g .z=(a z + b )(c z+ d )^{-1} , \quad g=\left( \begin{array}{c c} a & b \\ c & d \end{array} \right )\in SU(1,1),$$
where $g'$ is the complex Jacobian.
   Moreover, we have the following results:

$\bullet$ {
The discrete part of the spectrum of $\sH_\nu$ is not empty if and only if that $\bet >1$. It consists of the eigenvalues $\epsilon
_{m}^{\nu }$ given through \eqref{DiscSpec} and occurring with infinite multiplicities.
}

$\bullet$ {
Let $\nu$ be such that $\bet >1$. Then, for every fixed $m=0,1,2,\cdots ,\left[\nu-(1/2)\right]$, we have
\begin{enumerate}

\item[(i)] The family of functions given explicitly in terms of the Jacobi polynomials $P_{j}^{(\alpha,\beta)}(\cdot)$ as
\begin{align}\label{3.9}
\phi_{k}^{\nu,m}(z) =(-1)^{\min(m,k)} (1-|z|^2)^{-m} |z|^{|m-k|}
e^{-i(m-k)\arg z}P_{\min(m,k)}^{(|m-k|,2(\nu-m)-1) }(1-2|z|^{2})
\end{align}
constitutes an orthogonal basis of the $L^2$-eigenspace
\begin{eqnarray}
\mathcal{A}_{m}^{2,\nu }(\D):=\left\{ F:\D\longrightarrow %
\C,\quad F\in L^{2,\nu }(\D)\quad \mbox{ and }\quad \sH_{\nu}F=\epsilon _{m}^{\nu }F\right\} .  \label{3.7}
\end{eqnarray}

\item[(ii)] The square norm of $\phi_{k}^{\nu,m}$ in $L^{2,\nu }(\D)$ is given by
\begin{eqnarray}
\rho^{\nu,m}_{_k}=\left(\frac{\pi }{2(\nu-m)-1}\right) \frac{(\max(m,k))!\Gamma(2(\nu-m)+\min(m,k))}{(\min(m,k))!\Gamma(2(\nu-m)+\max(m,k))}.
\label{L2normb}
\end{eqnarray}
\end{enumerate}
}

\noindent Therefore the set of functions
\begin{align}
\Phi _{k}^{\nu ,m}:= \frac{\phi_{k}^{\nu,m}}{\sqrt{\rho^{\nu,m}_{_k}}};\qquad k=0,1,2,\cdots , \label{onbb}
\end{align}
constitute an orthonormal basis of $\mathcal{A}_{m}^{2,\nu }(\D)$. Moreover, using
 the identity \cite[p.63]{Szego}:
\begin{eqnarray}
\frac{\Gamma(m+1)}{\Gamma(m-s+1)}P_{m}^{(-s,\alpha)}(t)
=\frac{\Gamma(m+\alpha +1)}{\Gamma(m-s+\alpha+1)}\left(\frac{t-1}{2}\right)^{s}P_{m-s}^{(s,\alpha) }(t) , \quad 1\leq s\leq m,
\label{6.24}
\end{eqnarray}
which reads for  $s=m-k$, $t=1-2|z|^2$ and $\alpha=2(\nu-m)-1$ as
\begin{eqnarray}
P_{m}^{\left( k-m,\alpha \right) }\left( t \right)  =
(-1)^{m-k}\frac{ k!\Gamma \left( \bet-m\right)}{m!\Gamma\left( 2(\nu-m)+k\right) }
|z|^{2(m-k)} P_{k}^{\left( m-k,\alpha \right) }(t),  \label{6.25d}
\end{eqnarray}
we obtain
\begin{align}
(-1)^{k} \left(\frac{k! \Gamma(\bet -m)}{m! \Gamma(2(\nu-m)+k)} \right)^{\frac 12}
& \overline{z}^{m-k}P_{k}^{\left( m-k,\alpha \right) }(t) \label{corcor}\\
&=(-1)^{m}\left(\frac{m! \Gamma(2(\nu-m)+k) }{k! \Gamma(\bet -m) } \right)^{\frac 12}
z^{k-m}P_{m}^{\left( k-m,\alpha \right) }(t). \nonumber
\end{align}
Therefore, using \eqref{corcor} we check the following

$\bullet$ {
The function in \eqref{onbb} can be rewritten as
\begin{align}
\Phi_{k}^{\nu,m}(z)=(-1)^{k}&\left(\frac{2(\nu-m)-1}{\pi}\right)^{1/2}
   \left(\frac{k!\Gamma(2(\nu-m)+m)}{ m!\Gamma(2(\nu-m)+k)}\right)^{1/2} \label{Orthonbasis} \\
&\quad \times (1-|z|^{2})^{-m} \overline{z}^{m-k}P_{k}^{(m-k,2(\nu-m)-1)}(1-2|z|^{2}).  \nonumber
\end{align}
}

$\bullet$ {
The space $\mathcal{A}_{m}^{2,\nu }\left( \D\right) $ is a reproducing kernel Hilbert space.
Its $L^2$-eigenprojector kernel is given by
\begin{align}
K^{\nu}_m(z,w) &=\left(\frac{2(\nu-m)-1}{\pi}\right)
\left(1-z\bar{w}\right) ^{-\bet}
\left( \frac{|1-z\bar{w}|^{2}}{(1-|z|^{2})(1-|w|^{2})}\right)^{m} \label{rkB2} \\
&\qquad \qquad \qquad \times P_{m}^{(0,2(\nu-m)-1)}\left( 2\frac{(1-|z|^{2})(1-|w|^{2})}{|1-z\bar{w}|^{2}}-1\right) ,
\nonumber
\end{align}
with the diagonal function
\begin{eqnarray}
K^{\nu}_m(z,z)=\left(\frac{2(\nu-m)-1}{\pi }\right)(1-|z|^{2})^{-\bet},\quad z\in \D.  \label{3.12}
\end{eqnarray}
}

\begin{remark}\label{rem2.1}
In view of \eqref{3.9}, the $L^{2}$-eigenspace $\mathcal{A}_{0}^{2,\nu }(\D)$, corresponding to $m=0$ in \eqref{3.7} and associated to the bottom energy $\epsilon^\nu_0=0$, reduces further to the weighted Bergmann space $\mathcal{A}^{2,\nu }(\D)$ consisting of complex-valued holomorphic functions $F$ on $\D$ such that
\begin{eqnarray}
\int_{\D}|F(z)|^{2}(1-|z|^{2})^{\bet-2}d\mu (z)<+\infty. \label{3.13}
\end{eqnarray}
\end{remark}
\begin{remark}
The condition $\bet>1$ ensuring the existence of the eigenvalues \eqref{DiscSpec} should implies
that the magnetic field $B=d\theta_{\nu}=\bet\Omega (z)$, where $\Omega$ stands for the Khäler $2$-form on $\D$,
has to be strong enough to capture the particle in a closed orbit. If this condition is not fulfilled the motion will be
unbounded and the particle escapes to infinity.
\end{remark}

\section{Iwata's coherent states}

The first model of coherent states was the `nonspreading wavepacket' of the
harmonic oscillator, which have been constructed by Schrödinger \cite
{Schrodinger26}. In suitable units, wave functions of these states can be
written as
\begin{eqnarray}
\Phi _{\frak{z}}(\xi ):=\left\langle \xi \mid \frak{z}\right\rangle =\pi ^{-%
\frac{1}{2}}\exp \left( -\frac{1}{2}\xi ^{2}+\sqrt{2}\xi \frak{z}-\frac{1}{2}%
\frak{z}^{2}-\frac{1}{2}\left| \frak{z}\right| ^{2}\right) ,\quad \xi \in \R,  \label{2.1}
\end{eqnarray}
where $\frak{z}\in \Bbb{C}$ determines the mean values of coordinate $%
\widehat{x}$ and momentum $\widehat{p}$ according to $\left\langle \widehat{x%
}\right\rangle :=\left\langle \Phi _{\frak{z}},x\Phi _{\frak{z}%
}\right\rangle =\sqrt{2}\Re \frak{z}$ and $\left\langle \widehat{p}%
\right\rangle :=\left\langle \Phi _{\frak{z}},p\Phi _{\frak{z}}\right\rangle
=\sqrt{2}\Im \frak{z}.$ The variances $\sigma _{x}=\left\langle \widehat{x}%
^{2}\right\rangle -\left\langle \widehat{x}\right\rangle ^{2}=\frac{1}{2}$
and $\sigma _{p}=\left\langle \widehat{p}^{2}\right\rangle -\left\langle
\widehat{p}\right\rangle ^{2}=\frac{1}{2}$ have equal values, so their
product assumes the minimal value permitted by the Heisenberg uncertainty
relation. The coherent states $\Phi _{\frak{z}}$ have been also obtained by
Feymann \cite{Feynman51} and Glauber \cite{Glauber51} from the vacuum state $%
\mid 0\rangle $ by means of the unitary displacement operator $\exp \left(
\frak{z}A^{*}-\overline{\frak{z}}A\right) $ as
\begin{eqnarray}
\Phi _{\frak{z}}=\exp \left( \frak{z}A^{*}-\overline{\frak{z}}A\right) \mid
0\rangle ,  \label{2.2}
\end{eqnarray}
where $A$ and $A^{*}$ are respectively the annihilation and the creation operators defined by
\begin{eqnarray}
A=\frac{1}{\sqrt{2}}\left( \widehat{x}+i\widehat{p}\right) ,\qquad A^{*}=%
\frac{1}{\sqrt{2}}\left( \widehat{x}-i\widehat{p}\right)  \label{2.3}
\end{eqnarray}
It was Iwata \cite{Iwata} who used the well known expansion over the Fock
basis $|n\rangle $ to give an expression of $\Phi _{\frak{z}}$ as
\begin{eqnarray}
\Phi _{\frak{z}}=e^{-\frac{1}{2}\left| \frak{z}\right|
^{2}}\sum\limits_{n=0}^{+\infty }\frac{\frak{z}^{n}}{\sqrt{n!}} |n\rangle
.  \label{2.4}
\end{eqnarray}

Actually, various generalizations of coherent states are proposed.
Here, we shall focus on a generalization \textit{''à la Iwata''} of %
\eqref{2.4}. In the general setting, the procedure can be described as
follows. Let $(X,\mes)$ be a measure space and $\mathcal{A}^{2}\subset
L^{2}(X,\mes )$ be a closed subspace of infinite dimension. Let $\left\{
f_{k}\right\} _{k=1}^{\infty }$ be an orthogonal basis of $\mathcal{A}^{2}$
satisfying
\begin{eqnarray}
\omega \left( u\right) :=\sum_{k=1}^{\infty }\rho _{k}^{-1}\left|
f_{k}\left( u\right) \right| ^{2}<+\infty  \label{4.1}
\end{eqnarray}
for every $u\in X$, where $\rho _{k}:=\left\| f_{k}\right\| _{L^{2}(X)}^{2}$. Therefore the function
\begin{eqnarray}
K(u,v):=\sum_{k=1}^{\infty }\rho _{k}^{-1}f_{k}(u)\overline{f_{k}(v)},
\label{4.2}
\end{eqnarray}
defined on $X\times X$, is a reproducing kernel of the Hilbert space $%
\mathcal{A}^{2}$ so that we have $\omega \left( u\right) =K(u,u)$; $u\in X$.

\begin{definition}
For given infinite dimensional Hilbert space $\mathcal{H}$ with $\left\{
\psi _{k}\right\} _{k=1}^{\infty }$ as an orthonormal basis, the vectors $%
\left( \Psi _{u}\right) _{u\in X}$ defined by
\begin{eqnarray}
\Psi _{u}:=(\omega (u))^{-\frac{1}{2}}\sum_{k=1}^{\infty }\frac{f_{k}(u)}{%
\sqrt{\rho _{k}}}\psi_{_k}\quad \quad   \label{4.3}
\end{eqnarray}
will be called coherent states of Iwata type for the data of $(X;\mathcal{A}^2;\{f_{k}\})$ and $(\mathcal{H};\{\psi _{k}\})$.
\end{definition}
The choice of the Hilbert space $\mathcal{H}$ defines in fact a quantization of $%
X=\left\{ u\right\} $ by the coherent states $\Psi _{u}$, via the inclusion
map $u\rightarrow \Psi _{u}$ from $X$ into $\mathcal{H}$. Moreover, according to the fact that $\left\langle \Psi _{u},\Psi _{u}\right\rangle _{\mathcal{H}%
}=1 $, one can show the following

$\bullet$ {
The transform given by
\begin{eqnarray}
\W[f](u):=(\omega (u))^{\frac{1}{2}}\left\langle \Psi_{u},f\right\rangle _{\mathcal{H}} \label{cst}
\end{eqnarray}
defines an isometry from $\mathcal{H}$ into $\mathcal{A}^{2}$.
}

\noindent Thereby we have a resolution of the identity, i.e., the following
integral representation holds:
\begin{eqnarray}
f(\cdot )=\int_{X}\left\langle \Psi _{u},f\right\rangle _{\mathcal{H}}\Psi
_{u}(\cdot )\omega(u)\mes(u)  \label{resId}
\end{eqnarray}
for every $f\in \mathcal{H}$.

\begin{definition}
\label{cstDef} The transform $\W:\mathcal{H}\rightarrow \mathcal{A}^{2}\subset L^{2}(X,\mes )$ in \eqref{cst} will be called the coherent state
transform (CST) associated to the set of coherent states $\Psi _{u}$; $u\in X$.
\end{definition}

For an overview of all aspect of the theory of coherent states, we refer to the survey of V.V.
Dodonov \cite{Dodonov} or also to the recent book \cite{Gazeau} by J.P. Gazeau.

\section{Coherent states attached to Landau Levels $\epsilon _{m}^{\nu }$}

Now, we are in position to attach to each hyperbolic Landau level $\epsilon
_{m}^{\nu }$ in \eqref{DiscSpec} a set of generalized coherent
states according to formula \eqref{4.3}. Namely, we have
\begin{eqnarray}
\Psi_{\nu,m;z}:=\left( K^{\nu}_m(z,z) \right) ^{-\frac{1}{2}%
}\sum_{k=0}^{+\infty }\frac{\phi_{k}^{\nu,m}(z) }{\sqrt{\rho^{\nu,m}_{_k}}}\psi_{\nu,m;k}  \label{4.5}
\end{eqnarray}
with the following specifications:

\begin{itemize}
\item  $(X,\mes )=(\D,(1-|z|^2)^{\bet-2}d\mu)$.

\item  $\mathcal{A}^{2}=\mathcal{A}_{m}^{2,\nu }\left( \D\right) $
is the eigenspace in \eqref{GBS}.

\item  $K^{\nu}_m(z,z)$\textbf{\ }$=\pi ^{-1}\left( 2(\nu-m)-1\right)
(1-|z|^{2})^{-\bet }$ as in \eqref{3.12}.

\item  $f_k=\phi_{k}^{\nu,m}$ are the eigenfunctions given by \eqref{3.9}.

\item  $\rho^{\nu,m}_{_k}$\ being the square norm of $\Phi _{k}^{B,m}$
given in \eqref{L2normb}.

\item  $\mathcal{H}=L^{2}(\Bbb{R}_{+}^{*},\xi ^{-1}d\xi )$ is the Hilbert
space carrying the coherent states \eqref{4.5}.

\item  $\psi_k=\psi_{\nu,m;k}$, $k=0,1,2,\cdots ,$ the basis of $\mathcal{H}$ given by
\begin{eqnarray}
\psi_{\nu,m;k}(\xi ):=\left( \frac{k!}{\Gamma (2(\nu-m)+k)}%
\right) ^{\frac{1}{2}}\xi ^{\nu -m}e^{-\frac{1}{2}\xi }L_{k}^{\left(
2(\nu-m)-1\right) }(\xi ),\quad \xi >0.  \label{4.7} \qquad
\end{eqnarray}
\end{itemize}

In view of \eqref{4.5} and \eqref{Orthonbasis}, the coherent states belonging to the
Hilbert space $\mathcal{H}$ and corresponding to the eigenspace in \eqref{GBS} are defined by their
wave functions through the series expansion
\begin{align}
\Psi_{\nu,m;z}(\xi ):= &(1-|z|^{2})^{\nu -m}
\sum_{k=0}^{+\infty}(-1)^{k} \left( \frac{k!\Gamma (2(\nu-m)+m)}{m!\Gamma (2(\nu-m)+k)}\right) ^{1/2}  \label{CohSt2} \\
&\times \overline{z}^{m-k} P_{k}^{(m-k,2(\nu-m)-1)}(1-2|z|^{2})\psi_{\nu,m;k}(\xi ).  \nonumber
\end{align}
A closed form for \eqref{CohSt2} can be obtained in terms of Laguerre
polynomials as follows.

\begin{proposition}\label{Prop}
Let 
$\bet >1$ and $m=0,1,2,\cdots ,\left[ \nu -(1/2)\right] $.
 Then, the wave functions of the states in \eqref{CohSt2} read simply as
\begin{align}
\Psi_{\nu,m;z}(\xi )=(-1)^{m}\left( \frac{m!}{\Gamma (\bet -m)}%
\right) ^{\frac{1}{2}}& \frac{\left| 1-z\right| ^{2m}  }{\left( 1-z\right)^{\bet}}(1-|z|^{2})^{\nu -m}\xi ^{\nu -m}  \label{6.1} \\
& \times \exp \left( -\frac{\xi }{2}\frac{1+z}{1-z}\right) L_{m}^{2(\nu-m)-1}\left( \xi \frac{1-|z|^{2}}{|1-z|^{2}}\right) .  \nonumber
\end{align}
\end{proposition}

\noindent{\it Proof.} Set $\alpha =2(\nu-m)-1$ and $t=1-2|z|^2$. Then, the expression of $\Psi_{\nu,m;z}(\xi)$ in \eqref{CohSt2} reads as
\begin{align}
\Psi_{\nu,m;z}(\xi):=(1-|z|^2)^{\nu -m} \sum\limits_{k=0}^{+\infty}
                     (-1)^{k} \left(\frac{k! \Gamma(\alpha+1+m)}{m! \Gamma(\alpha+1+k)} \right)^{\frac 12}                       \overline{z}^{m-k} P_{k}^{(m-k,\alpha)}(t) \psi_{\nu,m;k}(\xi). \label{6.18cf}
\end{align}
By inserting the explicit expression of $\psi_{\nu,m;k}(\xi)$ given by \eqref{4.7} in \eqref{6.18cf}, we infers
\begin{eqnarray}
\Psi_{\nu,m;z}(\xi)
&=&  \left(\frac{\Gamma(\alpha+1+m)}{m!} \right)^{\frac 12} (1-|z|^2)^{\nu -m}\xi^{\nu -m}e^{-\frac{1}{2}\xi }\\
&\quad & \qquad \qquad \qquad \times\sum\limits_{k=0}^{+\infty } \frac{(-1)^{k}k!}{\Gamma(\alpha+1+k)}
\overline{z}^{m-k}P_{k}^{( m-k,\alpha) }(t)L_{k}^{(\alpha) }(\xi) \nonumber\\
&=&  \left(\frac{\Gamma(\alpha+1+m)}{m!} \right)^{\frac 12} (1-|z|^2)^{\nu -m}\xi^{\nu -m}e^{-\frac{1}{2}\xi }\\
&\quad & \qquad \qquad \qquad \times\sum\limits_{k=0}^{+\infty } \frac{k!}{\Gamma(\alpha+1+k)}
\overline{z}^{m-k}P_{k}^{\left(\alpha , m-k\right) }\left( -t\right)L_{k}^{(\alpha) }(\xi). \nonumber
\end{eqnarray}
The last equality is readily derived by means of the symmetry relation
\begin{eqnarray}
 P_{k}^{(a,b)}(t) =  (-1)^{k} P_{k}^{(b,a)}(-t).
\end{eqnarray}
In order to use the bilateral generating function  (\cite[p.213]{Rainville}):
\begin{align}\label{bgf}
\sum_{k=0}^{+\infty}\lambda^k{_2F_1}(- k, b; 1+\alpha; y)L_k^{(\alpha)}(\xi)
&= \frac{(1-\lambda)^{b-1-\alpha}}{(1-\lambda+y\lambda)^{b}} \exp\left({\frac{-\xi \lambda}{1-\lambda}}\right)\\
 &\qquad \quad \times{_1F_1}\left( b; 1+\alpha; \frac{\xi y\lambda}{(1-\lambda)(1-\lambda+y\lambda)}\right),\nonumber
\end{align}
involving a Laguerre polynomial and a terminating Gauss hypergeometric ${_2F_1}$-sum, we make appeal to the fact \cite[p.254]{Rainville}:
\begin{eqnarray}
P_{k}^{(\alpha,\eta )}(x) =\frac{\Gamma(1+\alpha+k)}{k!\Gamma(1+\alpha)}
\left(\frac{1+x}{2}\right)^{k}{_{2}F_{1}}\left( -k, -(\eta +k) ,1+\alpha ;\frac{x-1}{x+1}\right)
\end{eqnarray}
with $\eta = m-k$ and $x=-t=-1+2|z|^2$. Hence, we obtain
\begin{align}
\Psi_{\nu,m;z}(\xi)
=  \left(\frac{\Gamma(\alpha+1+m)}{m!} \right)^{\frac 12}& \frac{1}{\Gamma(\alpha+1)}
(1-|z|^2)^{\nu -m}\xi^{\nu -m}e^{-\frac{1}{2}\xi }\\
&\times\sum\limits_{k=0}^{+\infty }
 \overline{z}^{m}z^k{_{2}F_{1}}\left( -k, -m ;1+\alpha ;\frac{t+1}{t-1}\right)
L_{k}^{\left( \alpha\right) }(\xi). \nonumber
\end{align}
Thus, by applying \eqref{bgf} with  $\lambda=z$, $b =-m$ and $y= \frac{t+1}{t-1}=-\frac{1-|z|^2}{|z|^2}$, we check that
\begin{align}
\Psi_{\nu,m;z}(\xi)
&=  \left(\frac{\Gamma(\alpha+1+m)}{m!} \right)^{\frac 12} \frac{(-1)^m}{\Gamma(\alpha+1)}
\frac{|1-z|^{2m} }{(1- z)^{\bet}}(1-|z|^2)^{\nu -m} \label{psipsi}\\
& \qquad \times  \xi^{\nu -m} \exp\left(-\frac{\xi}2 \left(\frac{1+z}{1-z}\right)\right)
 {_1F_1}\left(-m ; \alpha+1; \frac{\xi (1-|z|^2)}{|1-z|^2}\right).\nonumber
 \end{align}
Finally, making use of \cite[p.103]{Szego}
\begin{eqnarray}
  {_1F_1}(-m; 1+\alpha; x) = \frac{m!\Gamma(1+\alpha)}{\Gamma(1+\alpha+m)} L^{(\alpha)}_m(x),
 \end{eqnarray}
with $x= \xi(1-|z|^2)/|1-z|^2$ yields
\begin{align}
\Psi_{\nu,m;z}(\xi)=&
 (-1)^m \left(\frac{m!}{\Gamma(\bet -m)}\right)^{\frac 12} \frac{|1-z|^{2m}}{(1- z)^{\bet}} (1-|z|^{2})^{\nu-m} \label{4.15}\\
&\qquad \times  \xi^{\nu -m} \exp\left({-\frac{\xi} 2 \left(\frac{1+ z}{1-z}\right)} \right) L^{(2(\nu-m)-1)}_m\left(\frac{\xi (1-|z|^2)}{|1-z|^2}\right).\nonumber
\end{align}
This completes the proof.\hfill $\Box$\\

According to Definition \ref{cstDef}, the coherent state transform associated with the
coherent states in \eqref{4.15} is the unitary map:
\begin{eqnarray}
   \W_{\nu ,m} &:& L^{2}(\Bbb{R}_{+}^{*},d\xi/\xi )\longrightarrow \mathcal{A}_{m}^{2,\nu }(\C)\\
                                            &\quad & \phi  \longmapsto  \W_{\nu ,m}\left[ \phi \right] (z):=\left( K^{\nu}_m(z,z)\right) ^{\frac{1}{2}}\left\langle \Psi_{\nu,m;z},\phi
\right\rangle _{L^{2}(\Bbb{R}_{+}^{*},\xi ^{-1}d\xi )}.  \label{BargTrans}
\end{eqnarray}
Explicitly, we have
\begin{eqnarray}
\W_{\nu ,m}\left[ \phi \right] (z) &=&\left( \frac{m!(2(\nu-m)-1)}{\pi \Gamma (\bet -m)}\right) ^{\frac{1}{2}}\left( \frac{1-|z|^{2}}{%
|1-z|^{2}}\right) ^{-m}(1-z)^{-\bet }  \label{G2ndBT} \\
&&\times \int_{0}^{+\infty }\xi ^{\nu -m}\exp \left( -\frac{\xi }{2}%
\left( \frac{1+z}{1-z}\right) \right) L_{m}^{2(\nu-m)-1}\left( \xi \frac{%
1-|z|^{2}}{\left| 1-z\right| ^{2}}\right) {\phi (\xi )}\frac{d\xi }{%
\xi }  \nonumber
\end{eqnarray}
thanks to Proposition \ref{Prop}. The assertion of Theorem \ref{Thm} follows then from the fact that the CST in  \eqref{cst} is an isometry.

\begin{definition}
The coherent state transform $\W_{\nu ,m}$
in \eqref{G2ndBT} will be called a generalized second Bargmann transform of index $m=0,1,2,\cdots,[\nu - (1/2)]$.
\end{definition}

\begin{remark}
For $m=0$, the above transform in \eqref{G2ndBT} reads simply
\begin{eqnarray}
\W_{\nu ,0}\left[ \phi \right] (z)=\left( \frac{{\bet-1}}{\pi
\Gamma (\bet )}\right) ^{\frac{1}{2}}(1-z)^{-\bet
}\int_{0}^{+\infty }\xi ^{\nu }\exp \left( -\frac{\xi }{2}\left(
\frac{1+z}{1-z}\right) \right) {\phi (\xi )}\frac{d\xi }{\xi }
\label{7.1}
\end{eqnarray}
and then reduces to the second Bargmann transform in \eqref{SBTM}.
\end{remark}

\begin{remark}
One can replace the space $\mathcal{H}$ by the weighted Bergman
 space in Remark \ref{rem2.1} to define a type of coherent states (see \cite{M1} for their series expansion).
The corresponding coherent state transform maps eigenstates of the first
hyperbolic Landau level $\epsilon^\nu_{0}$ into eigenstates corresponding
to $m^{th}$ level  $\epsilon^\nu_m$ as an integral transform  $\mathcal{A}_{0}^{2,\nu }(\D)\rightarrow
\mathcal{A}_{m}^{2,\nu }(\D)$. 
\end{remark}

\end{document}